# SECURING THE FUTURE OF HEALTHCARE: BUILDING A RESILIENT DEFENSE SYSTEM FOR PATIENT DATA PROTECTION

Ejiofor Oluomachi, Akinsola Ahmed

Department of Computer Science, Austin Peay State University,
Clarksville, USA.

## ABSTRACT

*The increasing importance of data in the healthcare sector has led to a rise in cybercrime targeting patient information. Data breaches pose significant financial and reputational risks to many healthcare organizations including clinics and hospitals. This study aims to propose the ideal approach to developing a defense system that ensures that patient data is protected from the insidious acts of healthcare data threat actors. Using a gradient-boosting classifier machine learning model, the study predicts the severity of healthcare data breaches. Secondary data was collected from the U.S. Department of Health and Human Services Portal with key indicators. Also, the study gathers key cyber-security data from Kaggle, which was utilized for the study. The findings revealed that hacking and IT incidents are the most common type of breaches in the healthcare industry, with network servers being targeted in most cases. The model evaluation showed that the gradient boosting algorithm performs well. Therefore, the study recommends that organizations implement comprehensive security protocols, particularly focusing on robust network security to protect servers.*

## 1. INTRODUCTION

In the digital era, the healthcare business is undergoing a dramatic transition as modern technologies are integrated. Electronic health records (EHRs), telemedicine, and health information systems are becoming essential components of modern healthcare delivery, promising more efficiency, better patient care, and better clinical results [1]. Greaves et al. describe digital health as a multidisciplinary domain that aspires to improve the efficiency of patient monitoring, diagnosis, management, prevention, rehabilitation, and long-term care provision[2]. According to the World Health Organization (2016), digital health is a fast-emerging medical area having a substantial impact on increasing healthcare quality and efficacy, decreasing healthcare costs for patients, and clinical research. But there are serious drawbacks to this digital change, cyberattacks targeting healthcare facilities are becoming prevalent and sophisticated, and as such protecting patient data has become a major problem. This fact emphasizes the necessity of having a strong defensive mechanism in place to protect private patient data from online attacks [3].

Moreover, cyber threat actors see patient data which includes financial, medical, and personal information as a very valuable resource. Data breaches have serious repercussions for the healthcare industry, including lost revenue, operational difficulties, jeopardized patient safety, and damaged public confidence in healthcare organizations [4]. High-profile events in recent times have brought attention to the vulnerability of healthcare systems and the pressing need for efficient cybersecurity solutions against ransomware attacks on hospitals and data breaches from health networks. These hacks could endanger lives by interfering with essential healthcare services and disclosing personal information to unscrupulous parties. Thus, for healthcare institutions, maintaining the security and integrity of patient data is not just a technological need but also a moral duty [5]. On the other hand, the state of healthcare cybersecurity today indicates a dynamic and intricate threat landscape. Cybercriminals are using more advanced strategies to take advantage of loopholes in healthcare systems. These strategies frequently use obsolete security standards, inadequate staff training, and noncompliance with regulations. Ransomware assaults, phishing schemes, and insider threats are examples of common cyber threats that provide different difficulties for patient data protection. Attacks using ransomware halt healthcare operations by encrypting important data and requesting large ransoms to unlock it. Conversely, phishing attacks can trick medical personnel into disclosing private information or allowing unapproved access to protected networks. Insider threats introduce hazards from within the company, whether intentional or unintentional, considerably complicating the security picture [3].

## 2. LITERATURE REVIEW

Organizations have undergone tremendous change because of the digital transformation process, which has also dramatically changed markets, relationships, user experiences, and cultural differences. The digital transformation process is being accelerated using emerging technologies like blockchain, big data, and artificial intelligence (AI). However, these technologies also pose significant new security concerns, underscoring the need for cybersecurity as a critical component of the health sector [6]. Cyber dangers, on the other hand, may be seen as a global issue that impacts all types of enterprises. Therefore, one of the most important security concerns facing the public and private sectors as well as individuals' private lives generally is how to respond to security crises [7].

### 2.1. Evolution of Cyber Threats in Healthcare

Technological advancements and the increasing sophistication of cybercriminals are closely reflected in the complicated story of how cyber risks have evolved in the healthcare industry. Healthcare companies were among the first industries to embrace extensive cybersecurity measures rather slowly, along with many other industries. Initially, network security linkages were the primary targets of basic cyber assaults. Cybercriminals trying to get sensitive personal and medical data found healthcare systems to be more appealing targets when they made the switch from paper-based records to electronic health records (EHRs) [6]. To increase effectiveness, accessibility, and patient care, the healthcare industry also started digitizing patient records in the late 1990s and early 2000s. Although advantageous, this change also brought up new risks. Early cyber threats frequently consisted of straightforward viruses and malware intended to exfiltrate data and maintain persistence. Most of these threats were opportunistic, going after any system with inadequate security. Healthcare organizations mostly took a reactive approach, concentrating on setting up firewalls and antivirus programs to fend against these intrusions [8].

Furthermore, the healthcare industry has integrated cutting-edge technology during the last ten years, which has further complicated the cybersecurity environment. While the use of artificial intelligence (AI), machine learning (ML), and Internet of Things (IoT) devices has improved patient care, it has also brought up new dangers. IoT devices frequently lack strong security measures, which makes them prime targets for hackers. Examples of these devices are wearable health monitoring and smart medical equipment. While improving diagnostic and therapeutic capacities, AI and ML can bring additional hazards if algorithms and underlying data are corrupted [9]. Additionally, software supply chain attacksare becoming another point of entry for cyber attackers as they lead to backdoor access, malware installation, application downtime,
and data leakage such as passwords or private information [11]. Zero-day exploits aim to take advantage of software flaws that the vendor is unaware of, leaving systems vulnerable until a fix is created. Conversely, supply chain attacks affect all users downstream by compromising hardware or software at the source. The potential scope and severity of such attacks were brought to light by the 2020 Solar Winds assault, which affected several industries, including the healthcare sector [10].

The cybersecurity strategy in the healthcare industry has had to change in response to these ever-changing threats, becoming more proactive and all-encompassing. Conventional fortifications centered around the perimeter are no longer adequate. A multi-layered security approach including encryption, secure access restrictions, intrusion detection and prevention systems, and continuous monitoring is becoming more and more common in healthcare institutions. The focus has switched to a comprehensive strategy that includes people, policies, procedures, and technology [1]. Defense measures now must include cybersecurity awareness training. Employees in the healthcare industry are frequently the first to detect and respond to questionable activity, which can greatly lower risk. Organizational security cultures are fostered by regular training and simulation exercises, such as phishing tests [4].

However, teamwork and information exchange have become more important. Cyber dangers are continually developing, and no one business can protect against them alone. Industry-wide efforts like the Health Information Sharing and Analysis Center (H-ISAC) make it easier for healthcare firms to share threat intelligence and best practices. This collaborative strategy improves collective defense capabilities and allows for a faster reaction to new threats [12].

## 2.2. Digital Resilience in Healthcare

The healthcare sector's digital transformation has increased its sensitivity to cyber-attacks, which are acknowledged as one of the most important societal and organizational dangers in terms of likelihood and effect. The healthcare sector's embrace of digital technology has been accelerated by substantial changes in the present scenario. Healthcare companies now acquire and keep a growing amount of sensitive data about persons, infrastructures, and supply networks. Any event that impacts those systems may have a substantial influence on the healthcare organization's strategy and operations, as well as the stakeholders [13].

However, digitization initiatives provide both a threat and an opportunity for healthcare resiliency. Jovanovic et al [14], observed that lower-level transformative change can foster higher-level resilience. This link between resilience and digital transformation is studied further in the sections that follow, using a cyber-security perspective. Based on a survey of the literature, several aspects emerge that characterize the idea of digital resilience in healthcare. To begin, writers differ in their assessment of healthcare system resilience as unique, i.e. a discrete area of resilience [15]. In this environment, several general traits and taxonomies appear. For example, the extant literature distinguishes between proactive and reactive measurements, or, more broadly, anticipatory and response-oriented elements of the idea. Moreover, there are several ways to characterize the traits of resilient systems. These include broader qualities like preparedness, robustness, and recovery/adaptation, as well as systemic functions like absorptive, adaptive, and transformative capacity [14].

Furthermore, output-based definitions of such systems emphasize community trust, safety, flexibility and continuity of service, and ownership in less abstract terms (Carthey et al., 2001). According to Blanchet et al. [12], another noteworthy development in healthcare is the categorization of "resilience" as a boundary term that bridges barriers between scientific and public/political discourse. This fits with a larger trend in cross-disciplinary work that Manyena[16] pointed out. This emphasizes how crucial it is to balance detail and abstraction when modeling healthcare resilience. The research emphasizes the role that culture and management play in fostering healthcare resilience, further connecting this concept to other organizational resilience domains [17]. These results are not supported by the participant selection process or the developing research design. Furthermore, digital resilience, also known as cyber resilience in the healthcare industry is viewed as an extension of higher order/institutional resilience, in keeping with its systemic approach [18].

## 2.3. Theoretical Review

### 2.3.1. Socio-Technical Systems Theory

The concept known as Socio-technological Systems Theory looks at the intricate interactions that occur between social and technological elements in an organization. Early in the 1950s, Eric Trist and Ken Bamforth put forth the notion [19]. According to this idea, the harmonious coexistence of people (social systems) and technology (technical systems) is necessary for good organizational outcomes. It highlights that to maximize overall performance, boost employee happiness, and increase operational efficiency, the social and technological subsystems should be co-planned rather than built separately [20].

Moreover, the idea has undergone modifications and has been used in several sectors, stressing cooperative design, inclusive decision-making, and the significance of customizing technology to suit human requirements and organizational settings. This holistic approach, which promotes the notion that technological advancements should strengthen rather than weaken the social dynamics of the workplace, is still relevant in modern organizational and information systems design. It is important to comprehend the social dimension, nevertheless. Understanding the duties, responsibilities, and conduct of healthcare workers, such as physicians, nurses, secretaries, and IT specialists, is necessary for this [21]. Comprehensive training programs that educate personnel on data protection best practices and increase awareness of cybersecurity dangers are essential components of a robust defensive system for patient data protection. Through this training, a security-conscious culture is fostered where each person is aware of their responsibility for protecting the privacy and integrity of patient data [22].

Considering the technological component also entails putting strong cybersecurity measures like intrusion detection systems, firewalls, and encryption into place. Updating and monitoring these technologies regularly is necessary to guard against developing cyber threats. To prevent interruptions in the provision of healthcare, these technological solutions should be developed and implemented in a user-friendly manner that smoothly integrates with current processes [23]. Moreover, the amalgamation of social and technological facets implies that cooperation between IT experts and healthcare professionals is imperative in the creation and execution of cybersecurity protocols. User input, for example, may help designers of security protocols make sure they are workable and don't interfere with the delivery of healthcare services. To ensure that their perspectives and experiences influence the creation of these systems, healthcare professionals should also be involved in the decision-making process when it comes to the adoption of new technologies [21].

## 2.4. Empirical Studies

Araujo et al. [24], comprehend cyber resilience in its numerous situations and aspects. To this goal, bibliographic research was conducted using indirect documentation in articles, books, and publications on the subject. The key phases of resilience were identified, and a study was conducted to see how these stages had developed over time. Finally, a new suggestion for standing for the phases of cyber resilience was offered, based on a compilation of recommendations from the full framework examined in this work. This assessment underlines the necessity of cyber resilience and comprehending the stages that define it, as well as the need for broader integration into companies in the most diversified sectors of activity management.

Garcia-Perez et al. [25] investigated basic components that contribute to the transformational, adaptive, and absorptive skills necessary for health systems' digital resilience. The study discovered that a balanced base of cyber security knowledge development, uncertainty management, and consideration for the sector's high levels of systemic and organizational interdependence is critical for its digital resilience and the long-term viability of its digital transformation initiatives.

He et al. [26], identify the most important cybersecurity challenges, the health sector's solutions, and areas that need to be improved to combat the recent rise in cyberattacks (such as ransomware and phishing campaigns), which attackers have exploited to target people and technology vulnerabilities brought about by changes made to working procedures in response to the COVID-19 pandemic. The study's findings highlighted four critical areas where the health sector's cybersecurity capability needs to be reinforced in addition to nine major cybersecurity problems and the eleven important solutions that healthcare companies adopted to solve these issues. The outcome also revealed that ransomware, malware, phishing, and distributed denial-of-service attacks were the most common and important hacking techniques used throughout the epidemic.

## 3. METHODOLOGY

To improve cyber health through the protection of patient data, the study extracted data from the U.S. Department of Health and Human Services Portal covering occurrences between 2021 and 2024. This data was collated to understand patterns and trends in cyber-attacks on healthcare data in the United States. This data contains relevant indicators, which include healthcare organizations, their locations, impacted individuals, the nature of the breach, and the breach's specific site. These incidents happened over various days, months, and years, revealing an uneven distribution of data over time. Additional data were obtained for variables such as detection method, breach severity, and days before detection.

Building a resilient defense system demands understanding the severity of the breaches to patient information. Hence, the target variable for the machine learning model is "severity of the data breaches", while other variables made up the features. The severity of the data breaches consists of three (3) categories, namely, hard, medium, small. Understanding the severity of a breach can help prioritize resources and tailor defensive strategies to prevent high-severity incidents. Data collected was extensively explored through Exploratory Data Analysis (EDA) in order to detect errors and missing values. Any irregularities identified were dealt with accordingly using the appropriate data cleaning techniques.

## 3.1. Missing Values

The variables with missing values were found in columns that are of string data type (object). The features are state (4), covered entity type (1), and location of breached information. The missing values were handled by deleting the rows using Pandas module in Python. More specifically, the dropna method was adopted.

Table 1: Variables Details

| Variables | Non-Null Count | Missing Values | Dtype |
| --- | --- | --- | --- |
| Name of Covered Entity | 809 non-null | | object |
| State | 805 non-null | | object |
| Covered Entity Type | 808 non-null | | object |
| Individuals Affected | 809 non-null | | int64 |
| Breach Submission Date | 809 non-null | | datetime64[ns] |
| Type of Breach | 809 non-null | | object |
| Location of Breached Information | 808 non-null | | object |
| Detection Method | 809 non-null | | object |
| Breach Severity | 809 non-null | | object |
| Days before detection | 809 non-null | | object |

## 3.2. Ensemble Methods

Gradient Boosting Classifier (GBC) is a powerful ensemble method ideal for enhancing patient data protection in cyber health. GBC builds a robust predictive model by iteratively combining multiple weak learners, typically decision trees. Each iteration focuses on correcting errors from previous models, improving accuracy and resilience to overfitting. The final model,

$$F_M(x) = F_0(x) + \sum_{m-1}^{M} \alpha_m h_m(x)$$

integrates all weak learners. GBC's high accuracy and ability to handle imbalanced data make it effective in detecting complex patterns and anomalies in patient data, ensuring robust cybersecurity measures in healthcare information systems.

## 4. RESULTS AND DISCUSSION

Table 2: Descriptive Analysis

| Variables | Individuals Affected | Days before Detection |
| --- | --- | --- |
| Count | 804 | 804.00 |
| Mean | 174926.25 | 29.78 |
| Std | 749149.96 | 12.04 |
| Min | 500.00 | 10.00 |
| 25% | 1379.50 | 19.00 |
| 50% | 6700.50 | 29.00 |
| 75% | 46285.25 | 41.00 |
| Max | 8952212.00 | 50.00 |

The summary statistics for the dataset, which includes "Individuals Affected" and "Days before Detection," reveal several key insights. On average, 174,926.85 individuals are affected per breach, with a significant standard deviation of 749,149.96, indicating considerable variability in the number of individuals affected across different breaches. The minimum number of individuals affected is 500, and the maximum is a staggering 8,952,212, highlighting the wide range of breach impacts. For the "Days before Detection," the average detection time is 29.78 days, with a standard deviation of 12.04 days, suggesting some variability in detection times. The detection time ranges from a minimum of 10 days to a maximum of 50 days.

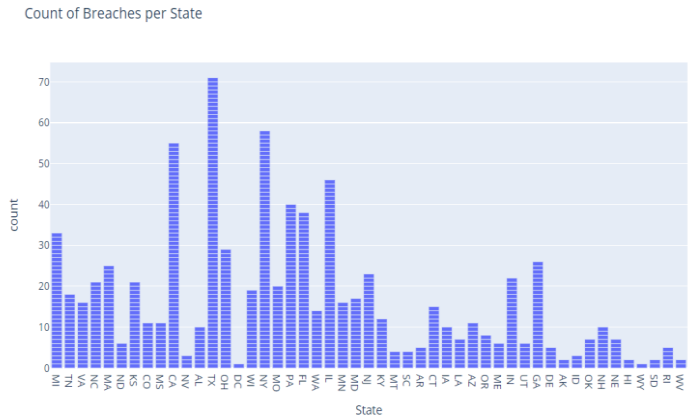

Figure 1: Count of Breaches per state

The summary statistics for the dataset indicating the number of breaches by state reveal the following insights: Among the top 5 most affected states is Texas (TX) which experienced the highest number of violations with 71 incidents, followed by New York (NY) with 58 incidents and California (CA) with 55 incidents. Illinois (IL) reported 46 breaches, while Pennsylvania (PA) had 40 breaches. This distribution suggests that Texas has the highest frequency of healthcare data breaches among the states listed, while Pennsylvania has the lowest within this subset.

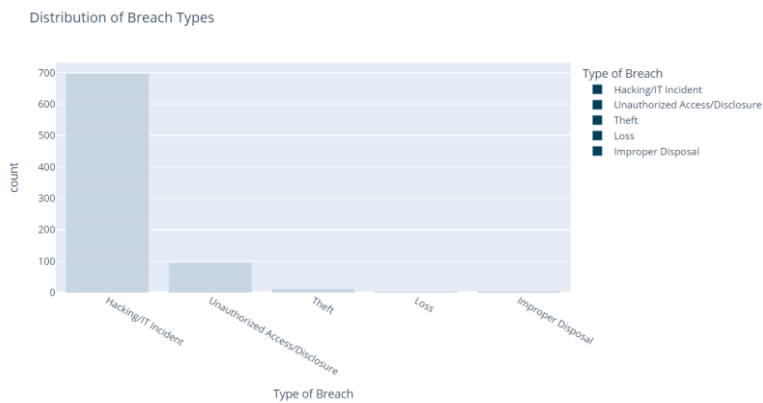

Figure 2: Type of Breach of Patient Data

The summary statistics for the types of data breaches reveal that Hacking/IT Incidents are the most prevalent, with 695 occurrences, making it the dominant category of breaches. Unauthorized Access/Disclosure follows, with 93 incidents, indicating a significantly lower but still notable frequency of breaches due to unauthorized activities. Theft accounts for 11 incidents, highlighting its lesser prevalence compared to the other types. Improper Disposal and Loss are the least common types of breaches, with only 3 and 2 occurrences, respectively. This distribution underscores the critical need for robust cybersecurity measures to address the predominant threat of hacking and IT incidents.

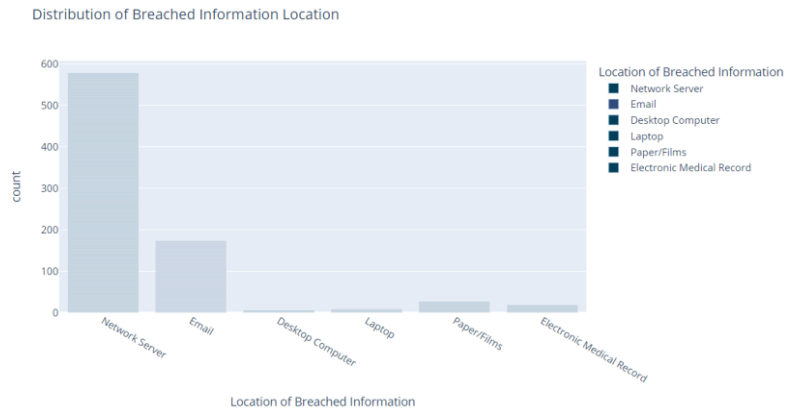

Figure 3: Location of Breached Information

The figure shows that network servers are the most common location for information breaches, with 577 incidents, highlighting the critical need for robust network security measures. Email breaches, totaling 172, emphasize vulnerabilities to phishing and unauthorized access. Physical documents (paper/films) experienced 26 breaches, indicating the ongoing importance of securing and properly disposing of paper records. Electronic medical records (EMRs) were breached 17 times, showing their susceptibility to cyber-attacks. Laptops and desktop computers had fewer breaches, with 7 and 5 incidents respectively, but still underscore the necessity of securing all devices containing sensitive information.

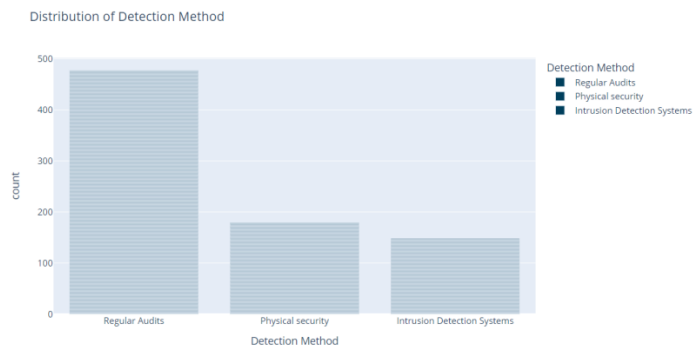

Figure 4: Method of Detecting Breaches

The above figure showed that most organizations detect breaches of patient health information through regular audits, while about 172 detect through physical security, and 148 discover through intrusion detection systems. This reflects the role of different methods in monitoring and alerting against unauthorized access and cyber threats. Together, these measures illustrate a comprehensive approach to safeguarding sensitive information through both proactive and reactive strategies.

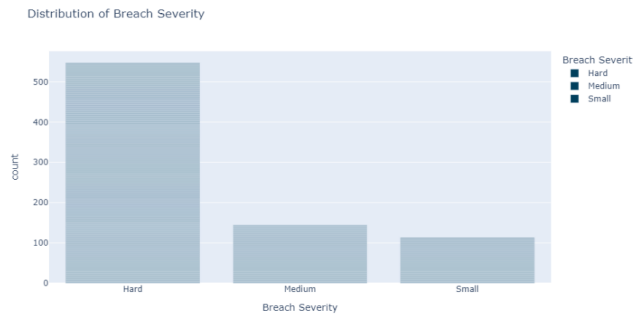

Figure 5: Distribution of Breach Severity

The figure presented categorizes breach severity into three distinct levels: Hard (547 incidents), Medium (144 incidents), and Small (113 incidents). This segmentation offers valuable insights into the distribution and implications of data breaches across different severity tiers. The predominance of breaches classified as Hard suggests a significant prevalence of severe security compromises or substantial data exposures.

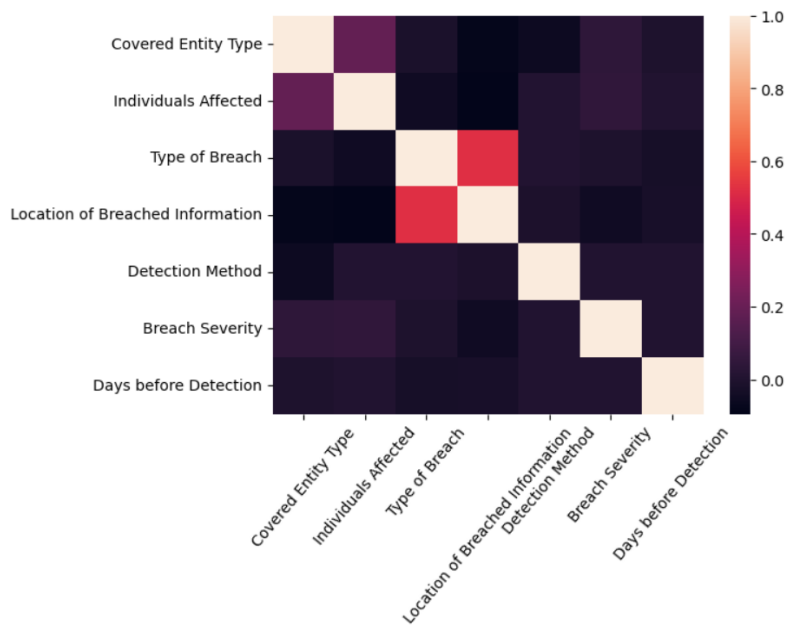

Figure 6: Heatmap showing the explanatory variables relationships

The color scale on the heatmap in Figure 6 indicates that most of the correlation coefficients fall below 0.4, except for the coefficients of correlation between the location of breached information and type of breach, which is 0.52. However, considering that the values fall below 0.8, it is concluded that none of the paired variables suffers from multicollinearity problems. Hence, the variables can be employed as features for the model.

### 4.1. Machine Learning Analysis

This study assessed the models using several key metrics: the F1 score, recall, accuracy, and precision. Accuracy is defined as the ratio of correctly predicted instances to the total number of instances in the dataset. However, relying solely on accuracy does not provide a complete evaluation of the model's performance. Therefore, it is crucial to consider the F1 score, recall, and precision in conjunction with accuracy.

The Gradient Boosting algorithm shows an accuracy score of 64%, and a precision score of 50.4%, indicating that the true positives predicted in the models are 50.4%. The models also demonstrated a recall score of 64%, indicating that the actual positive cases correctly identified by the gradient boosting algorithm are 64%. Lastly, the F1 score in the random forest is 55.5%.

## 4.2. The Confusion Matrix

The confusion matrix in Figure 7 presents the gradient boosting model's performance based on the classification of the target vector (Severity of the breach). Three categories are presented in the confusion matrix ("Hard", "Medium", "Small"). This relates to the ability of the gradient-boosting model to predict the severity of the breach of patient health information. The matrix displays the category distribution based on the prediction between the true label and the predicted label.

It was revealed that 153 instances of the "Hard" class were correctly classified, while 39 and 32 instances were incorrectly classified as "Medium" and "Small" respectively. From the second category, 2 instances of "Medium" were correctly classified, while 8 and 1 instances were incorrectly classified as "Hard" and "Small" respectively. Lastly, 0 instances of the "Small" category were correctly classified, while 7 instances were classified as "Hard".
Evaluation of the gradient boosting model showed that the model performed a little above average with regards to predicting the severe cases of patient data breaches, which implies that there is a need to increase the protection of health data, particularly patient information.

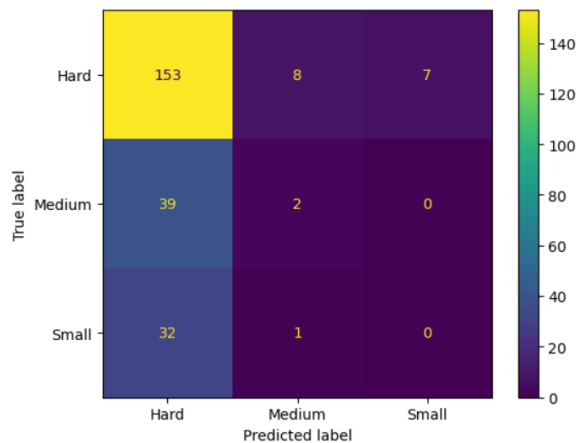

Figure 7: Confusion Matrix of the target classification

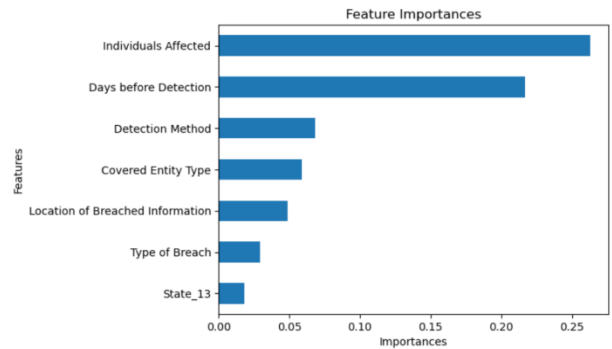

Figure 8: Feature Importances

Based on the importance of the features as extracted from the model, it was revealed that the most significant variable is the number of individuals affected, followed by the number of days before detection and method of detecting data breaches. The implication of this is that stakeholders in the healthcare industry showed place emphasis on the variables as presented in the figure above and this could help with building a resilient defense system for patient data protection.

## 5. DISCUSSION AND CONCLUSION

The analysis of healthcare data breaches highlights significant trends and the critical need for comprehensive security measures. On average, many individuals are affected per breach, with notable variability and detection times averaging close to a month. This underscores the importance of efficient breach detection mechanisms. Regionally, states like Texas, New York, and California exhibit higher incidences of breaches, pointing to specific vulnerabilities that may require targeted interventions. The prevalence of hacking and IT incidents as the primary breach type emphasizes the necessity for robust cybersecurity protocols. Network servers are the most common breach location, indicating that strengthening network security is paramount. Email breaches reflect vulnerabilities to phishing attacks, while the presence of breaches in physical documents, electronic medical records, laptops, and desktops highlights the need for comprehensive security that encompasses both digital and physical realms. Detection methods such as regular audits, physical security, and intrusion detection systems demonstrate the effectiveness of both proactive and reactive strategies in safeguarding sensitive information. The distribution of breach severity, with a predominance of severe incidents, suggests that substantial data exposures and security compromises are frequent, underscoring the urgency for heightened security measures and rapid response strategies to protect healthcare data effectively.

The machine learning algorithm (gradient boosting classifier) revealed the most significant features that are relevant to determining the severity of data breaches. The number of individuals affected, the number of days before detecting the breaches, and method of detecting it were found to be of high importance. The predictability of the model based on the evaluation method revealed that it performed above average, which implies that these features to a considerable extent can be used to determine severity of data breaches, thereby calling for taking necessary steps to building a formidable system to improve cyber health through patient data protection.

### 5.1. Recommendations

To address the critical need for enhanced healthcare data security, it is recommended that organizations implement comprehensive security protocols, particularly focusing on robust network security to protect servers. Regular audits and advanced intrusion detection systems should be prioritized for timely breach detection. Targeted interventions in high-incidence states like Texas, New York, and California can help mitigate regional vulnerabilities. Enhanced training programs to prevent phishing attacks and securing both digital and physical records are essential. Additionally, leveraging machine learning models to identify significant breach determinants can guide the development of more effective, predictive security measures to protect patient data and improve overall cyber health.

## AUTHORS

**Oluomachi Ejiofor** is a graduate student at Austin Peay State University studying Computer Science, specializing in Information Assurance and Security. She is passionate about cybersecurity, focusing on detecting threats, analyzingrisks, and developing secure software. With a proactive approach and a dedication to continuous learning, she aims to contribute to the cybersecurity field and help organizations stay protected in today's digital landscape

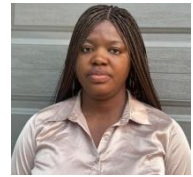

**Akinsola Ahmed** is a graduate student at Austin Peay State University in Clarksville, Tennessee, pursuing a Master of Science in Computer Science. He is passionate about information security and AI, and he believes these technologies have the potential to make the world a safer and more equitable place as such he's on a journey to leave an indelible mark in the world of technology, inspiring the next generation of innovators

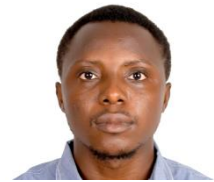